\documentclass[sigconf]{acmart}
\AtBeginDocument{%
  }

\acmDOI{}
\acmConference[UKAIRS '25]{Make sure to enter the correct
  conference title from your rights confirmation email}{September 08--09,
  2025}{Newcastle upon Tyne, UK}
\acmBooktitle{UKAIRS '25: UK AI Research Symposium,
  September 08--09, 2025, Newcastle upon Tyne, UK}
\acmISBN{}
\usepackage{xcolor, nccmath}

\begin{document}

%%
%% The "title" command has an optional parameter,
%% allowing the author to define a "short title" to be used in page headers.
\title{Treatment effect extrapolation in the presence of unmeasured confounding}

%%
%% The "author" command and its associated commands are used to define
%% the authors and their affiliations.
%% Of note is the shared affiliation of the first two authors, and the
%% "authornote" and "authornotemark" commands
%% used to denote shared contribution to the research.
\author{Stephanie Riley}
\orcid{0000-0002-7429-6655}
\affiliation{%
  \institution{University of Manchester, UK}
  %\city{Manchester}
  %\country{UK}
}
%\email{stephanie.riley-3@manchester.ac.uk}

\author{Ricardo Silva}
\affiliation{%
  \institution{University College London, UK}
  %\city{London}
  %\country{UK}
}
%\email{ricardo.silva@ucl.ac.uk}

\author{Matthew Sperrin}
\affiliation{%
  \institution{University of Manchester, UK}
  %\city{Manchester}
  %\country{UK}
}
%\email{matthew.sperrin@manchester.ac.uk}

%%
%% By default, the full list of authors will be used in the page
%% headers. Often, this list is too long, and will overlap
%% other information printed in the page headers. This command allows
%% the author to define a more concise list
%% of authors' names for this purpose.
\renewcommand{\shortauthors}{Riley et al.}

%%
%% The abstract is a short summary of the work to be presented in the
%% article.
\begin{abstract}
  While randomised controlled trials (RCTs) are the gold standard for estimating causal treatment effects, their limited sample sizes and restrictive criteria make it difficult to extrapolate to a broader population. Observational data, while larger, suffer from unmeasured confounding. Therefore, we can combine the strengths of both data sources for more accurate results.
  
  This work extends existing methods that use RCTs to debias conditional average treatment effects (CATEs) estimated in observational data by defining a deconfounding function. Our proposed approach borrows information from RCTs of multiple related treatments to improve the extrapolation of CATEs.

  Simulation results showed that, for non-linear deconfounding functions, using only one RCT poorly estimates the CATE outside of the support of that RCT. This is emphasised for smaller RCTs. Borrowing information from a second RCT provided more accurate estimates of the CATE outside of the support of both RCTs.
  
\end{abstract}

\maketitle

\vspace{0.5cm}

\section{Introduction}
Randomised controlled trials (RCTs) are considered the gold standard for estimating causal treatment effects. However, treatment effects estimated from RCTs do not generalise well to the wider population, as RCTs often have small sample sizes and restrictive inclusion criteria \cite{rothwell_2005}. In contrast, observational data are typically available for a given target population, with large sample size. Yet, since patients are not randomised to treatments in practice, there is unmeasured confounding. Therefore, we could utilise the complementary strengths of both observational and RCT data to extrapolate treatment effects \cite{degtiar_2023}. Existing work \cite{kallus_2018} uses a RCT to learn a ‘deconfounding’ function to debias conditional average treatment effects (CATEs) estimated using observational data. 

In this work we extend to the case where multiple RCTs are available for different but related treatments. Suppose the RCT corresponding to the treatment of interest is small with narrow inclusion criteria. However, we have information from other, larger RCTs, with wider inclusion criteria, looking at the effect of different treatments on the same outcome. We propose an approach that borrows information from the larger RCTs to extrapolate the CATE in the smaller RCT to a target population.

\section{Problem setup}
Suppose we have data from $k=1,...,K$ RCTs, each estimating the effect of some treatment $T_k$ on the outcome $Y$. The data also contain an effect modifier $X$. RCT $k$ thus comprises $\left\{ Y^k, T_k^k, X^k \right\}$ . We assume that the RCTs are unconfounded for their randomised treatments. We also have observational data, which contain unmeasured confounding, $U$, and observed data $\left\{ Y^0, T_1^0, \ldots, T_K^0, X^0 \right\}$ drawn randomly from the target population.

We are interested in extrapolating CATEs, $\tau_k(X) = \mathbb{E}[Y(T_k=1)|X] - \mathbb{E}[Y(T_k=0)|X]$, estimated in RCT $k$, to the target population. The extrapolated CATEs may be poorly estimated as they were learnt in a smaller subset of patients.

\begin{figure*}[t]
    \centering
    % \includesvg[width=\textwidth]{Indep_tau1_etaXsamplesize.svg}
    % \includegraphics[width=\textwidth]{Indep_tau1_etaXsamplesize.pdf}
    \includegraphics[width=0.93\textwidth]{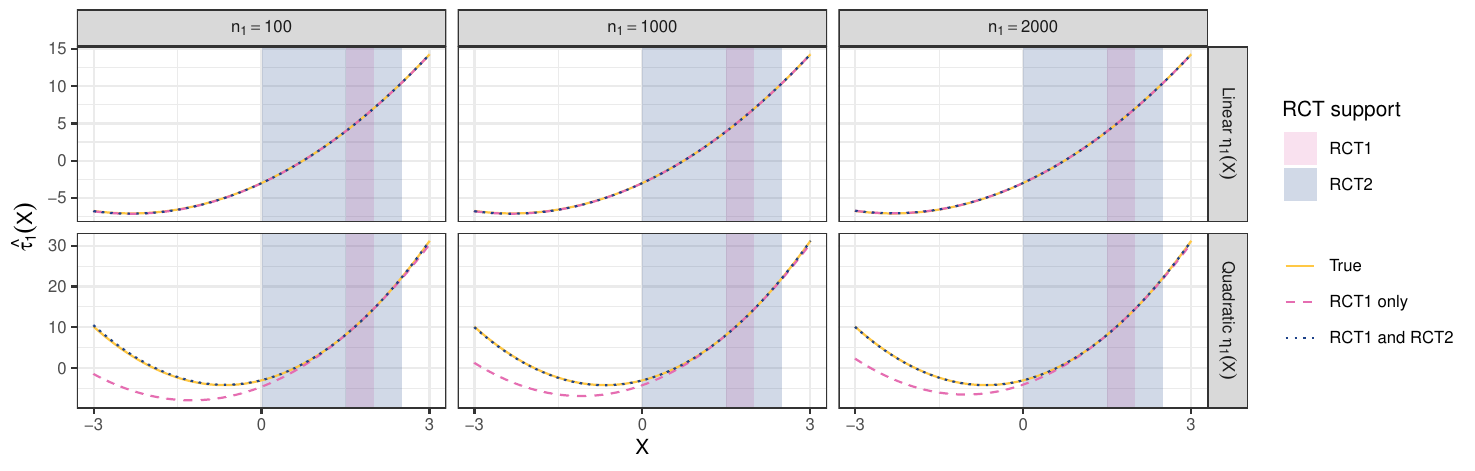}
    \vspace{-0.5cm}
    \caption{\small Summary of estimates of the conditional average treatment effect for treatment 1 ($\hat{\tau}_1(X)$). Panels are separated by the sample size of the smaller RCT (RCT1) and the true deconfounding function. The \textcolor[HTML]{e56db1}{pink} shaded region indicates the support of RCT1, and the \textcolor[HTML]{1d4289}{blue} shaded region the support of RCT2.}
    \label{fig:tau1}
\end{figure*}

\section{Proposed approach}
 Rather than extrapolate $\tau_k(X)$, we take inspiration from \cite{kallus_2018}, and estimate and extrapolate deconfounding functions,

\begin{align*}
    \eta_k(X) = \tau_k(X) - \omega_k(X),
\end{align*}

where $\omega_k(X)$ is a biased estimate of the CATE for treatment $k$, which we learn in the confounded observational data. 

We specify deconfounding functions,

\begin{align*}
    \eta_k(X) = \eta(X) + \epsilon_k(X),
\end{align*}

where $\eta(X)$ is shared across all of the studies, and $\epsilon_k(X)$ is the treatment-specific deviation from this shared unmeasured confounding. It seems reasonable to assume that there will be some shared unmeasured confounding, such as underlying frailty or disease progression.

Using the data from the RCTs to learn $\eta_k(X)$, we propose a hierarchical model of the form

\begin{align*}
    \mathbb{E} \left[ \eta_k(X) \right] = \boldsymbol{\beta}f(X) + \boldsymbol{\gamma}_k g(X)
\end{align*}

where $ \boldsymbol{\beta}$ is a vector of fixed effects, $\boldsymbol{\gamma}_k$ is a vector of random effects, and $f(X)$ and $g(X)$ describe the form of the shared and treatment-specific contributions to the deconfounding function, respectively. 

The deconfounding functions are then estimated using $\hat{\eta}_k(X) = \hat{\boldsymbol{\beta}}f(X) + \hat{\boldsymbol{\gamma}}_kg(X)$. Using the biased CATE estimated in the observational data, $\hat{\omega}_k(X)$, the debiased CATE can be estimated in the target population by $\hat{\tau}_k(X) = \hat{\omega}_k(X) + \hat{\eta}_k(X)$.

\section{Simulation}
Suppose we wish to estimate the CATE of treatment $T_1$ for some continuous outcome $Y$ in a target population, for which we have confounded data. We have a small RCT for the same treatment. We also have data from a larger RCT exploring the effect of another treatment ($T_2$) on the same outcome $Y$. We wish to utilise this larger trial to estimate deconfounding function $\eta_1(X)$.

We conducted a simulation study to evaluate our proposed approach and understand the impact of using multiple RCTs to extrapolate CATEs. Our estimands of interest were the estimate of the deconfounding function,  $\hat\eta_1(X)$, and corresponding CATE, $\hat\tau_1(X)$. 

We consider the cases where the true deconfounding functions are linear and quadratic. We also vary the sample size of the smaller RCT (for $T_1$) to understand how this impacts the estimation of the deconfounding function.

\subsection{Data generation}
\paragraph{Confounded data} 
We generated data for $n_0 = 50,000$ individuals in the observational data. An unmeasured confounding variable was generated from $U \sim \text{Bernoulli} (0.5)$ and a measured variable $X^0 \sim \text{Uniform} [-3, 3]$. Treatment indicators were independently generated for each of the $k = 1,2$ treatments: $T_1^0 | U = 1 \sim \text{Bernoulli} (0.7)$; $T_1^0 | U = 0 \sim \text{Bernoulli} (0.3)$;$~~T_2^0 | U = 1 \sim \text{Bernoulli} (0.25)$; and $T_2^0 | U = 0 \sim \text{Bernoulli} (0.75)$.

\paragraph{Deconfounding function}
Deconfounding functions, $\eta_k(X)$, were defined implicitly through the outcome variable $Y^k$. The outcome was generated from $Y^k \sim N(\mathbb{E}[Y|k], 1)$. For linear deconfounding functions, $\eta_1(X) = 2 - X$ and $\eta_2(X) = 1 - 2X$, 

{\small
\begin{align*}
    \mathbb{E}[Y|k=1] = 1 + 2X + 2T_1 + T_1X + 0.75T_1X^2 - 10UT_1 + 5UT_1X \\
    \mathbb{E}[Y|k=2] = 2 + X + T_2 + 1.5T_2X + T_2X^2 + 4UT_2 - 8UT_2X.
\end{align*}
}

For quadratic deconfounding functions, $\eta_1(X) = 2 - X -0.75X^2$ and $\eta_2(X) = 1 - 2X - 0.75X^2$,
{\small
\begin{align*}
    \mathbb{E}[Y|k=1] = 1 + &2X + 2T_1 + T_1X + 0.75T_1X^2 - \\
    &10UT_1 + 5UT_1X + 3.75UT_1X^2
\end{align*}

\begin{align*}
    \mathbb{E}[Y|k=2] = 2 + X + T_2 + 1.5T_2X + T_2X^2 + 4UT_2 - 8UT_2X - 3UT_2X^2.
\end{align*}
}

\paragraph{Unconfounded data}
We considered two RCTs: one small RCT looking at the CATE of $T_1$, and a larger RCT evaluating the CATE for $T_2$. The sample size was fixed for the second RCT, $n_2 = 5,000$. For the smaller RCT we varied the sample size, $n_1$, to include 100, 1,000, and 2,000. Treatment indicators in each RCT were generated from $T^k_k \sim \text{Bernoulli} (0.5)$. Measured variables were simulated from $X^1 \sim \text{Unif} [1.5, 2]$ and $X^2 \sim \text{Unif} [0, 2.5]$. Outcome variables were simulated as described above.

\subsection{Results}

We used two approaches to estimate the deconfounding function $\eta_1(X)$: using only the smaller RCT (RCT1 only), and using both RCTs (RCT1 and RCT2). When the true deconfounding function was linear, using either approach gave an accurate estimate of the CATE outside of the support of the RCTs (Figure \ref{fig:tau1}). This was consistent across all sample sizes that we considered for RCT1.

When the true deconfounding function was quadratic, using only one RCT poorly estimated the CATE in the target population. While estimation became more accurate as the sample size of that RCT increased, borrowing information from the larger RCT improved CATE estimation outside the support of both RCTs.

\section{Conclusion}
We propose an approach to correct for unmeasured confounding when estimating CATEs in observational data supported by RCTs. In simulations, we see that borrowing information from other RCTs allows us to more accurately extrapolate CATEs to target populations, particularly when the deconfounding function is non-linear.

%%
%% The acknowledgments section is defined using the "acks" environment
%% (and NOT an unnumbered section). This ensures the proper
%% identification of the section in the article metadata, and the
%% consistent spelling of the heading.
\begin{acks}
\small
We acknowledge support of the UKRI AI programme, and the Engineering and Physical Sciences Research Council, for CHAI - Causality in Healthcare AI Hub [grant number EP/Y028856/1].
\end{acks}

%%
%% The next two lines define the bibliography style to be used, and
%% the bibliography file.
\bibliographystyle{ACM-Reference-Format}
\bibliography{sample-base}

%%
%% If your work has an appendix, this is the place to put it.
% \appendix

% \section{Appendix}

\end{document}